# Twelve pure shear surface waves guided by bimaterial clamped/free boundaries in magneto-electro-elastic materials

Arman Melkumyan [*]

*Department of Mechanics, Yerevan State University, Alex Manoogyan Street 1, Yerevan 375025, Armenia*

*Abstract* — It is shown that surface waves with twelve different velocities in the cases of different magneto-electrical boundary conditions can be guided by the interface of two magneto-electro-elastic half-spaces. The plane boundary of one of the half-spaces is clamped while the plane boundary of the other one is free of stresses. The twelve velocities of propagation of these surface waves and the corresponding existence conditions are obtained in explicit forms. It is shown that the material coefficients of the half-space which has a clamped boundary have only quantitative influence on the surface waves and that the existence and absence of the possibility for the surface waves to be guided by the interface is determined by the material coefficients of the half-space which has a free boundary. The number of the possible different surface wave velocities decreases from 12 to 2 when the magneto-electro-elastic materials are changed to piezoelectric materials.

## I. Introduction

Surface acoustic wave (SAW) devices are widely used in numerous branches of science and technology and their investigation especially in the case of interconnected physical fields is an important and actively developing branch of research and applications. About 120 years ago the first type of SAW was described by Lord Rayleigh [1] in connection with the problem of earthquakes. Bleustein [2] and

---

[*] Tel. (+374 91) 482977.
  *E-mail address:* melk_arman@yahoo.com

Gulyaev [3] for the first time theoretically predicted that a pure shear SAW can be guided by the free surface of a piezoelectric half-space. Later Maerfeld and Tournois [4] investigated surface acoustic waves guided by the interface of two half-spaces and Danicki [5] described the SAW which can be guided by an embedded conducting plane in the electro-elastic materials in class 6 mm. Their fundamental results lay in the bases of further developments of acoustoelectronics and currently are cited in a great number of original papers.

Recent developments in physics and technology made possible to construct new materials called magneto-electro-elastic materials which demonstrate interconnection between magnetic, electric and elastic fields [6], [7]. When the electric field was connected with the elastic one (the piezoelectric materials) it brought up new and unexpected possibilities for science and technology. Connecting the magnetic field with the electric and elastic ones in magneto-electro-elastic materials suggests a range of new possibilities.

In this paper an investigation on the existence of pure shear surface acoustic waves guided by the interface of two transversely isotropic magneto-electro-elastic half-spaces in class 6 mm is investigated. Although much attention has been concentrated recently on magneto-electro-elastic materials and several dynamic problems have been solved by Alshits et al. [8], Hu and Li [9], Li [10], Chen et al. [11] etc., the investigation of surface acoustic waves in magneto-electro-elastic materials is currently an open actual subject. In this paper for the cases of different magneto-electrical boundary conditions in the plane boundaries of clamped and free magneto-electro-elastic half-spaces pure shear surface waves with 12 different velocities of propagation and the corresponding existence conditions are obtained. It is shown that the material coefficients of the half-space which has a clamped boundary have only a quantitative influence on the surface waves and that the existence and absence of the possibility for the surface acoustic waves to be guided by the interface is determined by the material coefficients of the half-space which has a free boundary. It is expected that these surface acoustic waves will have numerous applications in SAW devices and in further investigation of dynamic problems of magneto-electro-elastic materials. It is shown that the number of different pure shear surface acoustic wave velocities decreases from 12 to 2 when the magneto-electro-elastic materials are changed to piezoelectric materials.

## II. General equations and inequalities

Let $x_1$, $x_2$, $x_3$ denote rectangular Cartesian coordinates with $x_3$ oriented in the direction of the sixfold axis of the transversely isotropic magneto-electro-elastic materials in class 6 mm. Introducing the electric potential $\varphi$ and the magnetic potential $\phi$, so that

$$\mathbf{E}(x, y, t) = -\nabla \cdot \varphi(x, y, t), \quad \mathbf{H}(x, y, t) = -\nabla \cdot \phi(x, y, t), \tag{1}$$

where $\mathbf{E}$ is the electric field vector and $\mathbf{H}$ is the magnetic field vector, the five partial differential equations which govern the mechanical displacements $u_1$, $u_2$, $u_3$, and the potentials $\varphi$, $\phi$, reduce to two sets of equations when motions independent of the $x_3$ coordinate are considered. The equations of interest in the present paper are those governing the $u_3$ component of the displacement and the potentials $\varphi$, $\phi$, and can be written in the following form [10]:

$$c_{44}\nabla^2 u_3 + e_{15}\nabla^2 \varphi + q_{15}\nabla^2 \phi = \rho \ddot{u}_3,$$

$$e_{15}\nabla^2 u_3 - \varepsilon_{11}\nabla^2 \varphi - d_{11}\nabla^2 \phi = 0, \tag{2}$$

$$q_{15}\nabla^2 u_3 - d_{11}\nabla^2 \varphi - \mu_{11}\nabla^2 \phi = 0,$$

where $\nabla^2$ is the two-dimensional Laplacian operator, $\nabla^2 = \partial^2/\partial x_1^2 + \partial^2/\partial x_2^2$, $\rho$ is the mass density, $c_{44}$, $e_{15}$, $\varepsilon_{11}$, $q_{15}$, $d_{11}$ and $\mu_{11}$ are respectively the elastic modulus, piezoelectric coefficient, dielectric constant, piezomagnetic coefficient, magnetoelectric coefficient and magnetic permittivity constant, and the superposed dot indicates differentiation with respect to time. The constitutive equations which relate the stresses $T_{ij}$ ($i, j = 1, 2, 3$), the electric displacements $D_i$ ($i = 1, 2, 3$) and the magnetic induction $B_i$ ($i = 1, 2, 3$) to $u_3$, $\varphi$ and $\phi$ are

$$T_1 = T_2 = T_3 = T_{12} = 0, \quad D_3 = 0, \quad B_3 = 0,$$

$$T_{23} = c_{44}u_{3,2} + e_{15}\varphi_{,2} + q_{15}\phi_{,2}, \quad T_{13} = c_{44}u_{3,1} + e_{15}\varphi_{,1} + q_{15}\phi_{,1},$$

$$D_1 = e_{15}u_{3,1} - \varepsilon_{11}\varphi_{,1} - d_{11}\phi_{,1}, \quad D_2 = e_{15}u_{3,2} - \varepsilon_{11}\varphi_{,2} - d_{11}\phi_{,2}, \tag{3}$$

$$B_1 = q_{15}u_{3,1} - d_{11}\varphi_{,1} - \mu_{11}\phi_{,1}, \quad B_2 = q_{15}u_{3,2} - d_{11}\varphi_{,2} - \mu_{11}\phi_{,2}.$$

Solving Eqs. (2) for $\nabla^2 u_3$, $\nabla^2 \varphi$ and $\nabla^2 \phi$ and defining functions $\psi$ and $\chi$ by

$$\psi = \varphi - mu_3, \quad \chi = \phi - nu_3, \tag{4}$$

the solution of Eqs. (2) is reduced to the solution of

$$\nabla^2 u_3 = \frac{\rho}{\tilde{c}_{44}}\ddot{u}_3, \quad \nabla^2 \psi = 0, \quad \nabla^2 \chi = 0, \tag{5}$$

where

$$m = \frac{\mu_{11}e_{15} - d_{11}q_{15}}{\varepsilon_{11}\mu_{11} - d_{11}^2}, \quad n = \frac{\varepsilon_{11}q_{15} - d_{11}e_{15}}{\varepsilon_{11}\mu_{11} - d_{11}^2}, \tag{6}$$

and

$$\begin{aligned}
\tilde{c}_{44} &= c_{44} + \frac{\mu_{11}e_{15}^2 - 2d_{11}e_{15}q_{15} + \varepsilon_{11}q_{15}^2}{\varepsilon_{11}\mu_{11} - d_{11}^2} \\
&= \bar{c}_{44}^e + \frac{(\varepsilon_{11}q_{15} - d_{11}e_{15})^2}{\varepsilon_{11}(\varepsilon_{11}\mu_{11} - d_{11}^2)} \\
&= \bar{c}_{44}^m + \frac{(\mu_{11}e_{15} - d_{11}q_{15})^2}{\mu_{11}(\varepsilon_{11}\mu_{11} - d_{11}^2)}.
\end{aligned} \tag{7}$$

In Eqs. (5), (7) $\tilde{c}_{44}$ is the magneto-electro-elastically stiffened elastic constant, $\bar{c}_{44}^e = c_{44} + e_{15}^2/\varepsilon_{11}$ is the electro-elastically stiffened elastic constant and $\bar{c}_{44}^m = c_{44} + q_{15}^2/\mu_{11}$ is the magneto-elastically stiffened elastic constant.

Together with the electro-mechanical coupling coefficient $k_e^2 = e_{15}^2/(\varepsilon_{11}\bar{c}_{44}^e)$ and the magneto-mechanical coupling coefficient $k_m^2 = q_{15}^2/(\mu_{11}\bar{c}_{44}^m)$ introduce the magneto-electro-mechanical coupling coefficient

$$k_{em}^2 = \frac{\mu_{11}e_{15}^2 - 2d_{11}e_{15}q_{15} + \varepsilon_{11}q_{15}^2}{\tilde{c}_{44}\left(\varepsilon_{11}\mu_{11} - d_{11}^2\right)}$$

$$= \frac{e_{15}^2}{\tilde{c}_{44}\varepsilon_{11}} + \frac{\left(\varepsilon_{11}q_{15} - d_{11}e_{15}\right)^2}{\tilde{c}_{44}\varepsilon_{11}\left(\varepsilon_{11}\mu_{11} - d_{11}^2\right)}$$

$$= \frac{q_{15}^2}{\tilde{c}_{44}\mu_{11}} + \frac{\left(\mu_{11}e_{15} - d_{11}q_{15}\right)^2}{\tilde{c}_{44}\mu_{11}\left(\varepsilon_{11}\mu_{11} - d_{11}^2\right)}. \tag{8}$$

From Eqs. (6)-(8) follows that

$$\tilde{c}_{44} = c_{44}/\left(1 - k_{em}^2\right),$$

$$\left(\varepsilon_{11}q_{15} - d_{11}e_{15}\right)n = \tilde{c}_{44}\varepsilon_{11}k_{em}^2 - e_{15}^2,$$

$$\left(\mu_{11}e_{15} - d_{11}q_{15}\right)m = \tilde{c}_{44}\mu_{11}k_{em}^2 - q_{15}^2, \tag{9}$$

$$e_{15}m + q_{15}n = \tilde{c}_{44}k_{em}^2, \quad \varepsilon_{11}m + d_{11}n = e_{15}, \quad d_{11}m + \mu_{11}n = q_{15}.$$

Using the introduced functions $\psi$ and $\chi$ and the magneto-electro-elastically stiffened elastic constant, the constitutive Eqs. (3) can be written in the following form:

$$T_1 = T_2 = T_3 = T_{12} = 0, \quad D_3 = 0, \quad B_3 = 0,$$

$$T_{23} = \tilde{c}_{44}u_{3,2} + e_{15}\psi_{,2} + q_{15}\chi_{,2}, \quad T_{13} = \tilde{c}_{44}u_{3,1} + e_{15}\psi_{,1} + q_{15}\chi_{,1},$$

$$D_1 = -\varepsilon_{11}\psi_{,1} - d_{11}\chi_{,1}, \quad D_2 = -\varepsilon_{11}\psi_{,2} - d_{11}\chi_{,2}, \tag{10}$$

$$B_1 = -d_{11}\psi_{,1} - \mu_{11}\chi_{,1}, \quad B_2 = -d_{11}\psi_{,2} - \mu_{11}\chi_{,2}.$$

Using Eqs. (3) and the condition of the positiveness of energy one has that

$$c_{44} > 0, \quad \varepsilon_{11} > 0, \quad \mu_{11} > 0, \quad \varepsilon_{11}\mu_{11} - d_{11}^2 > 0. \tag{11}$$

From Eqs. (7), (8) and (11) one has that

$$\tilde{c}_{44} \geq c_{44}, \quad \tilde{c}_{44} \geq \overline{c}_{44}^e, \quad \tilde{c}_{44} \geq \overline{c}_{44}^m;$$

$$\tilde{c}_{44} = c_{44} \text{ if and only if } k_{em} = 0;$$

$$\tilde{c}_{44} = \overline{c}_{44}^e \text{ if and only if } n = 0; \tag{12}$$

$$\tilde{c}_{44} = \overline{c}_{44}^m \text{ if and only if } m = 0;$$

and

$$k_{em}^2 \geq \frac{e_{15}^2}{\tilde{c}_{44}\varepsilon_{11}}, \quad k_{em}^2 \geq \frac{q_{15}^2}{\tilde{c}_{44}\mu_{11}}, \quad 0 \leq k_{em} < 1;$$

$$k_{em}^2 = \frac{e_{15}^2}{\tilde{c}_{44}\varepsilon_{11}} \text{ if and only if } n = 0;$$

$$k_{em}^2 = \frac{q_{15}^2}{\tilde{c}_{44}\mu_{11}} \text{ if and only if } m = 0; \tag{13}$$

$$k_{em} = 0 \text{ if and only if } e_{15} = 0, \; q_{15} = 0.$$

If $d_{11} = 0$ then the expressions of the magneto-electro-elastically stiffened elastic constant and the magneto-electro-mechanical coupling coefficient are simplified:

$$\tilde{c}_{44} = c_{44} + \frac{e_{15}^2}{\varepsilon_{11}} + \frac{q_{15}^2}{\mu_{11}} = \overline{c}_{44}^e + \overline{c}_{44}^m - c_{44},$$

$$k_{em}^2 = \frac{e_{15}^2}{\tilde{c}_{44}\varepsilon_{11}} + \frac{q_{15}^2}{\tilde{c}_{44}\mu_{11}}. \tag{14}$$

Introduce short notations $e = e_{15}$, $\mu = \mu_{11}$, $d = d_{11}$, $\varepsilon = \varepsilon_{11}$, $q = q_{15}$, $c = c_{44}$, $\overline{c}^e = \overline{c}_{44}^e$, $\overline{c}^m = \overline{c}_{44}^m$, $\tilde{c} = \tilde{c}_{44}$, $w = u_3$, $T = T_{23}$, $D = D_2$, $B = B_2$ and use subscripts $A$ and $B$ to refer to the half-spaces $x_2 > 0$ and $x_2 < 0$, respectively.

Direct calculations show that if $\varepsilon_A$, $\mu_A$, $d_A$ and $\varepsilon_B$, $\mu_B$, $d_B$ satisfy the inequalities (11) then

a) $\varepsilon_A + \varepsilon_B$, $\mu_A + \mu_B$, $d_A + d_B$ also satisfy the inequalities (11), so that

$$\varepsilon_A + \varepsilon_B > 0, \; \mu_A + \mu_B > 0, \; (\varepsilon_A + \varepsilon_B)(\mu_A + \mu_B) - (d_A + d_B)^2 > 0; \tag{15}$$

b) the quadratic form

$$\frac{\mu_A e_A^2 - 2d_A e_A q_A + \varepsilon_A q_A^2}{\varepsilon_A \mu_A - d_A^2} - \frac{(\mu_A + \mu_B)e_A^2 - 2(d_A + d_B)e_A q_A + (\varepsilon_A + \varepsilon_B)q_A^2}{(\varepsilon_A + \varepsilon_B)(\mu_A + \mu_B) - (d_A + d_B)^2} = a_{11}e_A^2 - 2a_{12}e_A q_A + a_{22}q_A^2 \tag{16}$$

is a positive definite quadratic form, because

$$a_{11} = \frac{(\varepsilon_B \mu_B - d_B^2)\mu_A + (\varepsilon_B \mu_A^2 - 2d_B \mu_A d_A + \mu_B d_A^2)}{(\varepsilon_A \mu_A - d_A^2)\left[(\varepsilon_A + \varepsilon_B)(\mu_A + \mu_B) - (d_A + d_B)^2\right]} > 0, \tag{17}$$

$$a_{22} = \frac{\left(\varepsilon_B \mu_B - d_B^2\right)\varepsilon_A + \left(\mu_B \varepsilon_A^2 - 2d_B \varepsilon_A d_A + \varepsilon_B d_A^2\right)}{\left(\varepsilon_A \mu_A - d_A^2\right)\left[\left(\varepsilon_A + \varepsilon_B\right)\left(\mu_A + \mu_B\right) - \left(d_A + d_B\right)^2\right]} > 0, \tag{18}$$

$$a_{11}a_{22} - a_{12}^2 = \frac{\varepsilon_B \mu_B - d_B^2}{\left(\varepsilon_A \mu_A - d_A^2\right)\left[\left(\varepsilon_A + \varepsilon_B\right)\left(\mu_A + \mu_B\right) - \left(d_A + d_B\right)^2\right]} > 0. \tag{19}$$

### III. Surface waves

The mechanical conditions for the free and clamped surfaces of the magneto-electro-elastic materials which occupy the half-spaces $x_2 > 0$ and $x_2 < 0$ are

$$T_A = 0, \ w_B = 0 \ \text{on} \ x_2 = 0, \tag{20}$$

which must be satisfied together with the magneto-electrical conditions of an electrically closed ($\varphi = 0$) or electrically open ($D_2 = 0$) surface and magnetically closed ($B_2 = 0$) or magnetically open ($\phi = 0$) surface. The realization of each of these boundary conditions is described in [8].

The conditions at infinity require that

$$w_A, \ \varphi_A, \ \phi_A \to 0 \ \text{as} \ x_2 \to \infty,$$

$$w_B, \ \varphi_B, \ \phi_B \to 0 \ \text{as} \ x_2 \to -\infty. \tag{21}$$

Consider the possibility of a solution of Eq. (5) of the form of a surface waves propagating in the positive direction of the $x_1 = 0$ axis:

$$w_A = w_{0A} \exp(-\xi_A x_2) \exp\left[i(\xi_1 x_1 - \omega t)\right],$$

$$\psi_A = \psi_{0A} \exp(-\xi_1 x_2) \exp\left[i(\xi_1 x_1 - \omega t)\right], \tag{22}$$

$$\chi_A = \chi_{0A} \exp(-\xi_1 x_2) \exp\left[i(\xi_1 x_1 - \omega t)\right],$$

in the half-space $x_2 > 0$ and

$$w_B = w_{0B} \exp(\xi_B x_2) \exp\left[i(\xi_1 x_1 - \omega t)\right],$$

$$\psi_B = \psi_{0B} \exp(\xi_1 x_2) \exp[i(\xi_1 x_1 - \omega t)], \tag{23}$$

$$\chi_B = \chi_{0B} \exp(\xi_1 x_2) \exp[i(\xi_1 x_1 - \omega t)],$$

in the half-space $x_2 < 0$.

From Eqs. (20), (23) follows that $w_{0B} = 0$, so that $w_B$ is identically equal to zero. Then Eqs. (22)-(23) satisfy the conditions (21) if $\xi_1 > 0$, $\xi_A > 0$. The second and the third of Eqs. (5) are identically satisfied and the first of Eqs. (5) requires

$$\tilde{c}_A(\xi_1^2 - \xi_A^2) = \rho_A \omega^2. \tag{24}$$

Now the mechanical boundary conditions (20) together with different magneto-electrical boundary conditions on $x_2 = 0$ must be satisfied. Consider the following cases of magneto-electrical boundary conditions on $x_2 = 0$:

1a. $D_A = D_B = 0$, $\phi_A = \phi_B = 0$;

1b. $D_A = 0$, $\varphi_B = 0$, $\phi_A = \phi_B = 0$;

1c. $D_A = 0$, $\varphi_B = 0$, $B_B = 0$, $\phi_A = 0$;

1d. $D_A = D_B = 0$, $B_B = 0$, $\phi_A = 0$;

2a. $\varphi_A = \varphi_B = 0$, $B_A = B_B = 0$;

2b. $\varphi_A = \varphi_B = 0$, $B_A = 0$, $\phi_B = 0$;

2c. $\varphi_A = 0$, $D_B = 0$, $B_A = 0$, $\phi_B = 0$;

2d. $\varphi_A = 0$, $D_B = 0$, $B_A = B_B = 0$;

3a. $\varphi_A = \varphi_B = 0$, $\phi_A = \phi_B = 0$;

3b. $\varphi_A = 0$, $D_B = 0$, $\phi_A = \phi_B = 0$;

3c. $\varphi_A = \varphi_B = 0$, $B_B = 0$, $\phi_A = 0$;

3d. $\varphi_A = 0$, $D_B = 0$, $B_B = 0$, $\phi_A = 0$; \hfill (25)

4. $D_A = D_B = 0$, $B_A = B_B$, $\phi_A = \phi_B$;

5. $D_A = D_B$, $\varphi_A = \varphi_B$, $B_A = B_B = 0$;

6. $D_A = D_B$, $\varphi_A = \varphi_B$, $B_A = B_B$, $\phi_A = \phi_B$;

7. $D_A = D_B$, $\varphi_A = \varphi_B$, $\phi_A = \phi_B = 0$;

8. $\varphi_A = \varphi_B = 0$, $B_A = B_B$, $\phi_A = \phi_B$;

9. $D_A = 0$, $\varphi_B = 0$, $B_A = B_B$, $\phi_A = \phi_B$;

10. $D_A = D_B$, $\varphi_A = \varphi_B$, $B_A = 0$, $\phi_B = 0$;

11. $D_A = D_B$, $\varphi_A = \varphi_B$, $B_B = 0$, $\phi_A = 0$;

12. $\varphi_A = 0$, $D_B = 0$, $B_A = B_B$, $\phi_A = \phi_B$;

13a. $D_A = D_B = 0$, $B_A = B_B = 0$;

13b. $D_A = 0$, $\varphi_B = 0$, $B_A = B_B = 0$;

13c. $D_A = D_B = 0$, $B_A = 0$, $\phi_B = 0$;

13d. $D_A = 0$, $\varphi_B = 0$, $B_A = 0$, $\phi_B = 0$.

Each of the 25 groups of conditions in Eq. (25) together with Eqs. (20), (22)-(23) leads to a system of six homogeneous algebraic equations for $w_{0A}$, $\psi_{0A}$, $\chi_{0A}$, $w_{0B}$, $\psi_{0B}$, $\chi_{0B}$, the existence of a nonzero solution of which requires that the determinant of that system be equal to zero. This condition for the determinant and Eqs. (21), (24) determine the surface wave velocities $V_s = \omega/\xi_1$ and the existence conditions.

In the case of 1a) of Eqs. (25) this procedure leads to a surface wave with the following velocity

$$V_{s1}^2 = \frac{\tilde{c}_A}{\rho_A}\left(1-\gamma_1^2\right), \ \gamma_1 = k_{emA}^2 - \frac{e_A^2}{\tilde{c}_A \varepsilon_A}. \tag{26}$$

The same velocity is obtained in the cases 1b), 1c) and 1d). Each of the cases 2a), 2b), 2c) and 2d) leads to a surface wave with the velocity

$$V_{s2}^2 = \frac{\tilde{c}_A}{\rho_A}\left(1-\gamma_2^2\right), \ \gamma_2 = k_{emA}^2 - \frac{q_A^2}{\tilde{c}_A \mu_A}, \tag{27}$$

and each of the cases 3a), 3b), 3c) and 3d) leads to a surface wave with the velocity

$$V_{s3}^2 = \frac{\tilde{c}_A}{\rho_A}(1-\gamma_3^2), \quad \gamma_3 = k_{emA}^2. \tag{28}$$

The cases 13a), 13b), 13c) and 13d) do not lead to any surface wave. Each of the cases 4) to 12) leads to its own single surface wave and the corresponding surface wave velocities are $V_{si}^2 = \frac{\tilde{c}_A}{\rho_A}(1-\gamma_i^2)$ ($i=4,5,\ldots,12$) where

$$\gamma_4 = \frac{\varepsilon_A(\varepsilon_B\mu_B - d_B^2)}{\varepsilon_B(\varepsilon_A\mu_A - d_A^2) + \varepsilon_A(\varepsilon_B\mu_B - d_B^2)}\left[k_{emA}^2 - \frac{e_A^2}{\tilde{c}_A\varepsilon_A}\right];$$

$$\gamma_5 = \frac{\mu_A(\varepsilon_B\mu_B - d_B^2)}{\mu_B(\varepsilon_A\mu_A - d_A^2) + \mu_A(\varepsilon_B\mu_B - d_B^2)}\left[k_{emA}^2 - \frac{q_A^2}{\tilde{c}_A\mu_A}\right];$$

$$\gamma_6 = k_{emA}^2 - \frac{(\mu_A + \mu_B)e_A^2 - 2(d_A + d_B)e_A q_A + (\varepsilon_A + \varepsilon_B)q_A^2}{\tilde{c}_A\left[(\varepsilon_A + \varepsilon_B)(\mu_A + \mu_B) - (d_A + d_B)^2\right]};$$

$$\gamma_7 = k_{emA}^2 - \frac{e_A^2}{\tilde{c}_A(\varepsilon_A + \varepsilon_B)};$$

$$\gamma_8 = k_{emA}^2 - \frac{q_A^2}{\tilde{c}_A(\mu_A + \mu_B)}; \tag{29}$$

$$\gamma_9 = \frac{\varepsilon_A\mu_B}{\varepsilon_A\mu_B + (\varepsilon_A\mu_A - d_A^2)}\left[k_{emA}^2 - \frac{e_A^2}{\tilde{c}_A\varepsilon_A}\right];$$

$$\gamma_{10} = \frac{\varepsilon_B\mu_A}{\varepsilon_B\mu_A + (\varepsilon_A\mu_A - d_A^2)}\left[k_{emA}^2 - \frac{q_A^2}{\tilde{c}_A\mu_A}\right];$$

$$\gamma_{11} = k_{emA}^2 - \frac{\varepsilon_A\mu_B}{\varepsilon_A\mu_B + (\varepsilon_B\mu_B - d_B^2)}\frac{e_A^2}{\tilde{c}_A\varepsilon_A};$$

$$\gamma_{12} = k_{emA}^2 - \frac{\varepsilon_B\mu_A}{\varepsilon_B\mu_A + (\varepsilon_B\mu_B - d_B^2)}\frac{q_A^2}{\tilde{c}_A\mu_A}.$$

From Eqs. (11)-(13), (15)-(19) and (26)-(29) follows that $0 \le \gamma_i < 1$, so that $V_{si} > 0$ ($i=1,2,\ldots,12$). The Eqs. (21) and (24) require that $V_{si} < \tilde{c}_A/\rho_A$ ($i=1,2,\ldots,12$) which leads to the following existence conditions for the surface acoustic waves:

$n_A \neq 0$ for existence of the surface waves with velocities $V_{s1}$, $V_{s4}$, $V_{s9}$;

$m_A \neq 0$ for existence of the surface waves with velocities $V_{s2}$, $V_{s5}$, $V_{s10}$;

$k_{emA} \neq 0$ for existence of the surface waves with velocities $V_{s3}$, $V_{s6}$, $V_{s7}$, $V_{s8}$, $V_{s11}$, $V_{s12}$.

These existence conditions show that the material coefficients of the half-space $B$ ($x_2 < 0$) which has a clamped boundary have only quantitative influence on the surface acoustic waves and that the existence and absence of the possibility for the surface waves to be guided by the interface is determined by the material coefficients of the half-space $A$ ($x_2 > 0$) which has a free boundary.

If the magneto-electro-elastic materials degenerate to piezoelectric materials, so that $q_A \to 0$, $d_A \to 0$, $q_B \to 0$, $d_B \to 0$, the surface waves that have velocities $V_{s1}$, $V_{s4}$, $V_{s9}$ disappear and

$$V_{s2}^2, V_{s3}^2, V_{s8}^2, V_{s12}^2 \to \frac{\overline{c}_A^e}{\rho_A}\left(1 - k_{eA}^4\right); \tag{30}$$

$$V_{s5}^2, V_{s6}^2, V_{s7}^2, V_{s10}^2, V_{s11}^2 \to \frac{\overline{c}_A^e}{\rho_A}\left(1 - \left(\frac{\varepsilon_B}{\varepsilon_A + \varepsilon_B}\right)^2 k_{eA}^4\right), \tag{31}$$

so that the number of different surface wave velocities decreases from 12 to 2.

### IV. Conclusions

It is shown that new twelve pure shear surface acoustic waves in the cases of different magneto-electrical boundary conditions can be guided by the interface of two magneto-electro-elastic half-spaces one of which has a clamped boundary and the other one has a boundary which is free of stresses. The velocities of propagation of each of these surface waves and the corresponding existence conditions are obtained in explicit exact forms. It is shown that the material coefficients of the half-space which has a clamped boundary have only quantitative influence on the surface waves and that the existence and absence of the possibility for the surface waves to be guided by the interface is determined by the material coefficients of the half-space which has a free boundary. The number of the possible different pure shear surface acoustic wave velocities decreases from 12 to 2 if the magneto-electro-elastic materials are changed to piezoelectric materials.